\documentclass[final,3p,times,twocolumn]{elsarticle}

\usepackage{amssymb}

\journal{Physica A}

\begin{document}

\begin{frontmatter}



\title{Opinion formation driven by PageRank node influence on directed networks}


\author[a1,a2]{Young-Ho Eom}
\author[a2]{Dima L. Shepelyansky}
\address[a1]{IMT Institute for Advanced Studies Lucca,
Piazza San Francesco 19, Lucca 55100, Italy}
\address[a2]{{\it Laboratoire de Physique Th\'eorique du CNRS,
IRSAMC, Universit\'e de Toulouse, UPS, F-31062 Toulouse, France}}

\begin{abstract}
We study a two states opinion formation model driven by PageRank
node influence and report an extensive numerical study on how
PageRank affects collective opinion formations in large-scale
empirical directed networks. In our model the opinion of a node
can be updated by the sum of its neighbor nodes' opinions weighted
by the node influence of the neighbor nodes at each step. We
consider PageRank probability and its sublinear power as node
influence measures and investigate evolution of opinion under
various conditions. First, we observe that all networks reach
steady state opinion after a certain relaxation time. This time
scale is decreasing with the heterogeneity of node influence in
the networks. Second, we find that our model shows consensus and
non-consensus behavior in steady state depending on types of
networks: Web graph, citation network of physics articles, and
LiveJournal social network show non-consensus behavior while
Wikipedia article network shows consensus behavior. Third, we find
that a more heterogeneous influence distribution leads to a more
uniform opinion state in the cases of Web graph, Wikipedia, and
Livejournal. However, the opposite behavior is observed in the
citation network. Finally we identify that a small number of
influential nodes can impose their own opinion on significant
fraction of other nodes in all considered networks. Our study
shows that the effects of heterogeneity of node influence on
opinion formation can be significant and suggests further
investigations on the interplay between node influence and
collective opinion in networks.
\end{abstract}


\end{frontmatter}


\section{Introduction}
\label{}
Each individual has her/his own opinion about political, social,
and economical issues based on her/his own belief, information,
and perspective. Individuals also exchange, discuss, and reconcile
their opinions with others through social contacts or networks.
Through these interactions, collective opinions emerge from our
society. The recent advent of social media such as Twitter or
Facebook accelerates the emergence of collective opinions on
global scale. Understanding how collective opinions are formed on
various types of social networks has critical importance in the
era of information technology.

Statistical physics community has provided quantitative tools to
reveal the underlying mechanisms that govern the collective
opinion formation through social
interactions~\cite{Castellano2009}. Various opinion formation
models (see Refs. \cite{Castellano2009,Xia2011} for details) on
networks including voter
models~\cite{Galam1986,Galam2008,Sood2005,Suchecki2005}, majority
rule model~\cite{Galam2002},  bounded confidence
model~\cite{Deffuant2000}, and Sznajd
model~\cite{Sznajd-Weron2000} were suggested and extensively
studied. These models have given us analysis tools of how network
structure affects opinion dynamics and have provided us
mathematical understanding of collective opinion formation.

In order to expand our understanding of collective opinion
formation on networks further we can consider the following two
directions. First we can consider opinion formation on real social
networks rather than on artifact network models such as regular
lattices or small-world networks which are mainly considered in
previous studies~\cite{Castellano2009,Xia2011} and far from real
networks. Second, in most of real situations, there are opinion
leaders or elites who have strong influence and lead collective
opinions in social systems. The roles of these leaders or elites
on opinion formation is still elusive. In short, it is necessary
to understand how heterogeneous individual influence affects on
collective opinion formation on real networks.

In a recent study~\cite{Kandiah2012}, PageRank is proposed as a
node influence measure in an opinion formation model on
large-scale real networks such as Web graphs and social media
including LiveJournal and Twitter. The PageRank opinion formation
(PROF) model, introduced in~\cite{Kandiah2012}, takes into account
a node influence in the process of opinion formation. In the PROF
model, the opinion of a node is updated by the weighted sum of
neighbor nodes' opinions and the weight of the neighbor nodes are
given by their PageRank (see the next section for details). It is
found that a group of top influential elites in the networks
(i.e., nodes with high PageRank) can impose their own opinion on a
significant fraction of the considered
networks~\cite{Kandiah2012}. The PROF model is also considered on
Ulam networks~\cite{Chakhmakhchyan2013}, generated by the
intermittency map and the Chirikov typical map, showing a similar
behavior with the case of World Wide Web (WWW).

In the present work we consider how heterogeneous node influence
affects the collective opinion formation using the modified
PageRank opinion formation (PROF) model to go beyond previous
works~\cite{Kandiah2012,Chakhmakhchyan2013}. Our goal is to
examine how the PROF model behaves on real directed networks if we
adjust the heterogeneity of node influence (i.e., the PageRank of
nodes). The original PROF model considered only linear case of
PageRank as a node influence, it is necessary to consider opinion
formation driven by node influence under more general conditions.
To do this we modified the PROF model considering sublinear
PageRank of nodes such that the influence of node $i$ is given by
${P_i}^{g}$ where $P_i$ is the PageRank of node $i$ and $0\leq
g\leq 1$. Extensive numerical study of the model shows various
features of considered opinion formation. First we observed that
all networks reach a steady state opinion and the relaxation time
to this state is decreasing with the heterogeneity of node
influence in the networks. Second we found our model shows
consensus and non-consensus behavior in steady state depending on
types of networks: Web graph, citation network of physics
articles, and LiveJournal social network show non-consensus
behavior while Wikipedia article network shows consensus behavior.
Third we found that the more heterogeneous distribution of node
influence the network has (i.e., higher $g$), the more uniform
opinion state we can observe in Web graph, Wikipedia, and
Livejournal. However, in the citation network, the more
heterogeneous distribution of node influence leads to the less
uniform opinion. Finally we observed that a small number of
influential nodes can impose their own opinion on significant
fraction of other nodes in all considered networks.

The paper is organized as follows. The modified PROF model is
described in Section 2. The description of considered empirical
directed networks is given in Section 3. The extensive numerical
studies on empirical networks are presented in Section 4. A
discussion of the result is given in Section 5.

\section{Opinion formation by the modified PROF model}

We consider a directed network $G(N,L)$ with $N$
nodes and nodes in the network are connected by $L$ directed
links. Based on the network structure, the PageRank probability
$P_i(t)$ of node $i$ at iteration time $t$ is given by

\begin{equation}
P_i(t)=(1-\alpha)/N + \alpha\sum_{j} A_{ij}P_j(t-1)/k_{out}(j),
\label{eq:PageRank}
\end{equation}
where $A_{ij}$ is the adjacency matrix of the network $G$ and
$A_{ij}=1$ if there is a directed link from node $i$ to $j$,
$k_{out}(j)$ is the out-degree of node $j$ (i.e., number of
out-links from node $j$), and $\alpha$ is the damping
factor~\cite{Meyer2006}. In this study, we used the conventional
value $\alpha=0.85$~\cite{Meyer2006}. We take the stationary state
$P(i)$ of $P(i,t)$ as the PageRank of node $i$.

PageRank is a widely used node centrality to quantify influence of
nodes in a given directed network. Originally PageRank was
introduced for Google web search engine to rank web pages in World
Wide Web based on the idea of academic citations~\cite{Brin1998}.
Currently PageRank is used to rank nodes in various types of
directed networks including citation networks of scientific
papers~\cite{Chen2007,Frahm2014}, social network
services~\cite{Kwak2010}, world trade network~\cite{Ermann2011},
biological systems~\cite{Kandiah2013},
Wikipedia~\cite{Eom2013EPJB,Eom2013PLOS,Eom2015},
scientists~\cite{Radicchi2009}, and tennis
players~\cite{Radicchi2011}.

In this work each node $i$ has a binary opinion $\sigma_i \in
\{-1,+1\}$ and has PageRank $P_i$ as a node influence based on
network structure and Eq.~(\ref{eq:PageRank}). At each opinion
update, a node $i$ is randomly chosen and its opinion is updated
considering its neighbor nodes' opinions. Each time step consists
of $N$ updates. Thus one time step corresponds to one opinion
update for each node on average. The opinion updating rule
considers node influence of each neighbor node. Adopted from the
original PageRank opinion formation (PROF)
model~\cite{Kandiah2012,Chakhmakhchyan2013}, the update rule
reads: if the following function $H(i)$ for the chosen node $i$ is
positive, then $\sigma_i = +1$ otherwise $\sigma_i = -1$. The
function $H(i)$ is given by:

\begin{equation}
H(i) = a\sum_{j\in \Lambda_{i,in}} \sigma_j{P_j}^g + b\sum_{j\in
\Lambda_{i,out}} \sigma_j{P_j}^g, a+b=1
\end{equation}
where $\Lambda_{i,in}$ is the group of in-neighbor nodes of node
$i$ (i.e., the nodes have out-links to node $i$) and
$\Lambda_{i,out}$ is the group of out-neighbor nodes of node $i$
(i.e., the nodes have out-links from node $i$), respectively. The
parameter $g$ quantifies the heterogeneity of node influence. If
$g=0.$ then every node in the network has same node influence. If
$g=1.0$ then every node in the network can influence other nodes'
opinion as much as its PageRank and thus this case is reduced to
the original PROF model~\cite{Kandiah2012}. Thus, $H(i)$ is the
weighted summation of opinions of node $i$'s neighbor nodes. In
this study we use $a=b=0.5$ for simplicity of analysis.

\section{Empirical networks}
\label{}
We consider the following four empirical directed networks. (1)
\emph{Web graph}:  we consider Web graph of University of
Cambridge~\cite{Frahm2011,WebData}; here each node corresponds to a
Web page and a link is hyper-link between the Web pages in the
domain of University of Cambridge. (2) \emph{Citation network}: we
consider Physical Review citation network~\cite{Frahm2014}; here a
node corresponds to an article published in Physical Review
journal of American Physical Society from 1897 to 2009 and the
links correspond to the citation relations between the articles.
(3) \emph{Wikipedia}: we consider the network of articles in
French Wikipedia~\cite{Eom2015}; the nodes correspond to articles
in French Wikipedia (fr.wikipedia.org) and the links are the
inter-articles hyper-links between the articles. (4)
\emph{LiveJournal}: we consider the social network of
LiveJounral (livejournal.com) users; here the nodes are users of
LiveJournal and the links are social relationship between the
users; a more detail information on the network data are given
in~\cite{Kurucz2008}.

Statistical properties of the considered empirical networks are
represented in Table~\ref{table:1}. It is notable that unlike
typical networks such as regular lattices or small-world networks
considered in opinion formation models, all considered networks in
this work have complex structural properties including broad
degree distributions and broad distribution of
PageRank~\cite{Kandiah2012,Frahm2014,Eom2015,Frahm2011}.

\begin{table}[!ht]
  \caption{Basic statistics of empirical directed networks,
  $N$ gives the total number of nodes and $L$ gives the total number of links.}
\label{table:1}
\begin{center}
\begin{tabular}{c|c|c}
 \hline
Network & $N$ & $L $ \\ \hline
Web graph & 212710 & 1831542 \\
Citation & 463349 & 4690897  \\
Wikipedia & 1352825 & 34431943 \\
LiveJournal & 3577166 & 44913072 \\
 \hline
\end{tabular}
\end{center}
\end{table}

\section{Results}

\begin{figure*}[ht]
\begin{center}
\includegraphics[width=100mm,angle=-90]{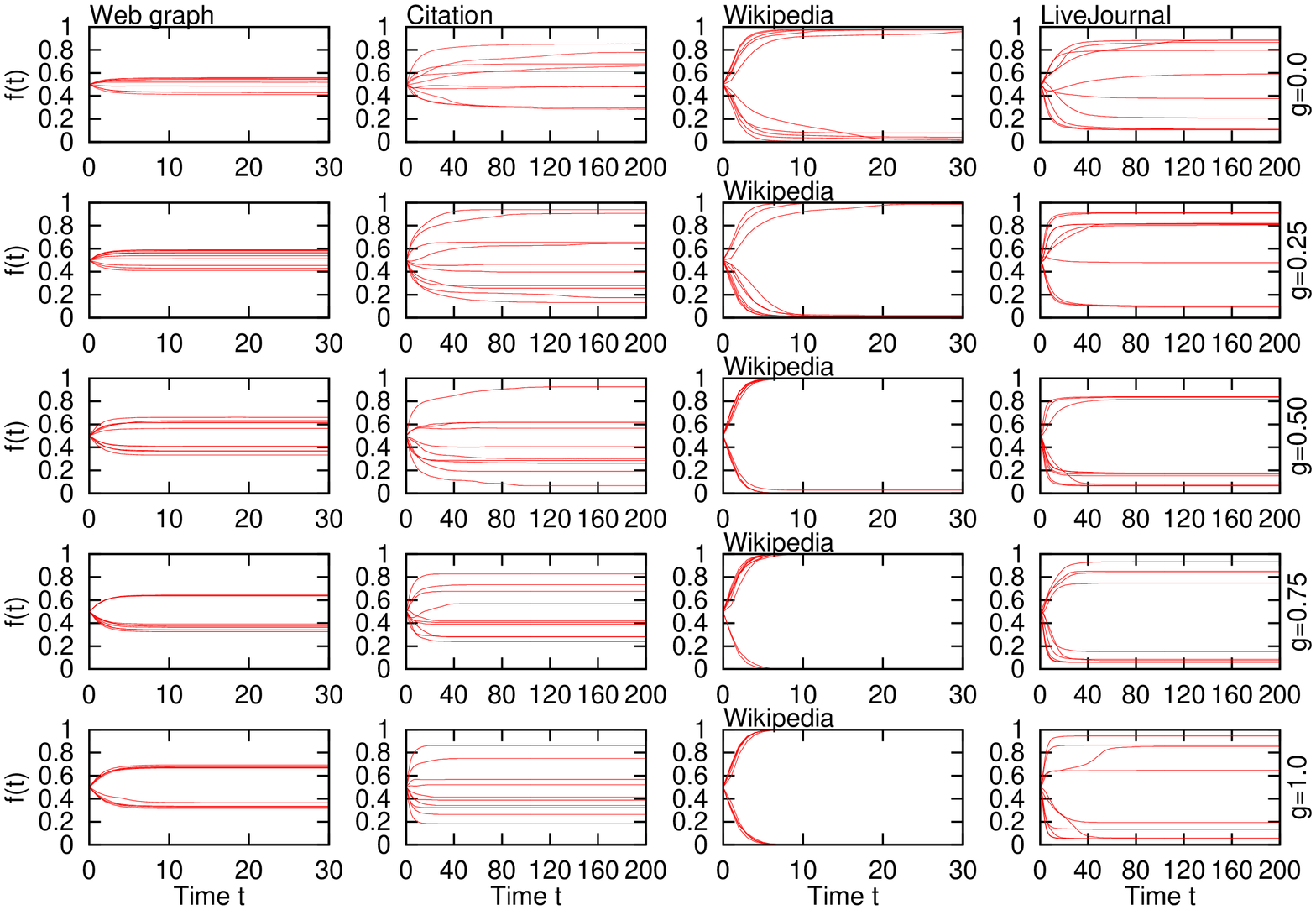}
\caption{Evolution of the fractions of $(+1)$ opinion $f(t)$ in
time $t$. Here $10$ realizations per each network and each value
of $g$ are represented. Each column corresponds to the network and
each row corresponds to $g$.} \label{fig:1}
\end{center}
\end{figure*}

With the modified PROF model on described empirical networks, we
investigate dynamics of collective opinion formation. First we
consider evolution of the fractions of $(+1)$ opinion, $f(t,+1)$,
by time $t$ to investigate whether considered networks can reach
the steady state or not and whether they reach consensus opinion
or not if the networks can reach the steady state. For simplicity,
we represent $f(t)=f(t,+1)$. By definition, we can consider the
fraction of $(-1)$ opinion $f(t,-1)=1-f(t)$ easily. Starting with
same initial fraction of two opinions (i.e.,
$f(0,+1)=f(0,-1)=0.5$), we numerically investigate how fractions
of each opinion state evolve by time t. As shown in
Fig.~\ref{fig:1}, all considered networks have reached the steady
states. Sub-figures located in the bottom row of Fig.~\ref{fig:1}
represent the evolution of the fraction of $(+1)$ opinion nodes
$f(t)$ along with time $t$ and $g=1$ ($10$ realizations for each
network). For Wikipedia case (the third column of
Fig.~\ref{fig:1}), we can observe ``consensus" behavior (i.e.,
most of nodes have single major opinion whether $(+1)$ or $(-1)$).
However, we observed that Web graph (the first column of
Fig.~\ref{fig:1}), Citation network (the second column of
Fig.~\ref{fig:1}), and LiveJournal social network (the fourth
column of Fig.~\ref{fig:1}) show non-consensus behavior (i.e., two
finite values of opinion co-exist in the steady states). Here we
define that if a given network have reached either $f_s>0.95$ or
$f_s<0.05$, the network shows consensus behavior where $f_s$ is
the fraction of $(+1)$ opinion in the steady state. We find that
Web graph and Wikipedia relax to the steady state (either
consensus or non-consensus) in short time ($t<30$) as shown in
Fig.~\ref{fig:1} while more longer times ($t>40$) are necessary to
reach the steady states in cases of Citation and LiveJournal
networks. Sub-linear $g$ values cases (figures from the first to
fourth row) show similar behaviors of reaching steady state with
the linear cases. But it is notable that for Web graph and
Wikipedia, the differences between each steady state fractions of
$(+1)$ opinions are bigger with growing $g$. We can consider this
observation as a sign of growing polarization of steady state
opinion. However, other networks give no clear signs. A further
more quantitative analysis for these gaps between the fraction of
steady state opinions are required.

To quantify the effects of $g$ value on the relaxation time to the steady state
of the collective opinion, first we define $\langle f(t)
\rangle_{10}$ as an average fraction of $(+)$ state for $10$
consecutive time steps from time $t$ to $t+9$ as following.

\begin{equation}
\langle f(t) \rangle_{10} = \frac{1}{10}\sum_{t}^{t+9}{f(t)}
\end{equation}

\begin{figure*}[ht]
\begin{center}
\includegraphics[width=74mm,angle=-90]{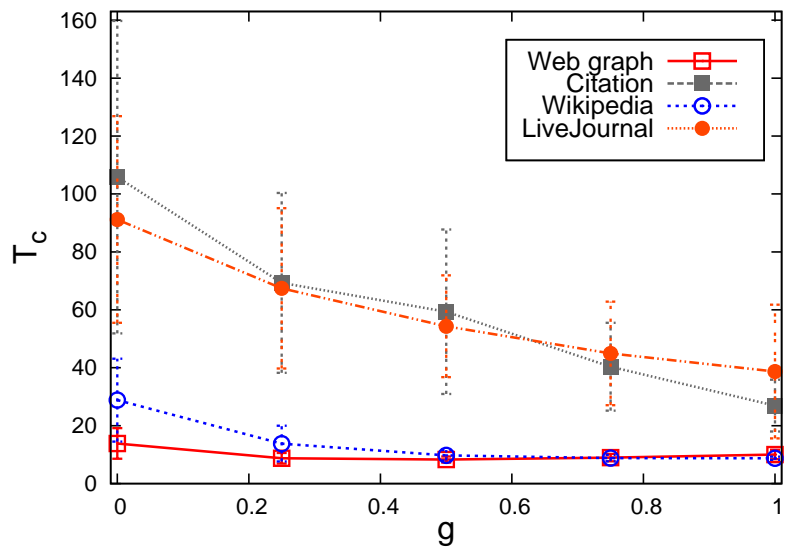}
\caption{Dependence of the relaxation time $T_c$ to a steady state
on the influence exponent $g$ for considered
networks.} \label{fig:2}
\end{center}
\end{figure*}

We define time $T_c$ of reaching the steady state for each network
such that the standard deviation $\sigma(10)$ of above ten
consecutive fraction $f(t)$ of $(+1)$ opinion nodes from time
$t=T_c$ to $t=T_c+9$ is less than $0.0002$. (i.e.,
$\sigma(10)<0.0002$). Fig.~\ref{fig:2} represents the relation
between steady state relaxation time $T_c$ and the influence
exponent $g$. We can observe a clear tendency that bigger $g$
(more heterogeneous influence the network has) leads to shorter
time to reach the steady states for all networks. As Fig.1.
implies, Web graph and Wikipedia have shorter relaxation times
$T_c<30$ for various $g$ while Citation and LiveJournal networks
have significantly longer $40<T_c<110$ and effects of $g$
variation are more pronounced.

\begin{figure*}[ht]
\begin{center}
\includegraphics[width=106mm,angle=-90]{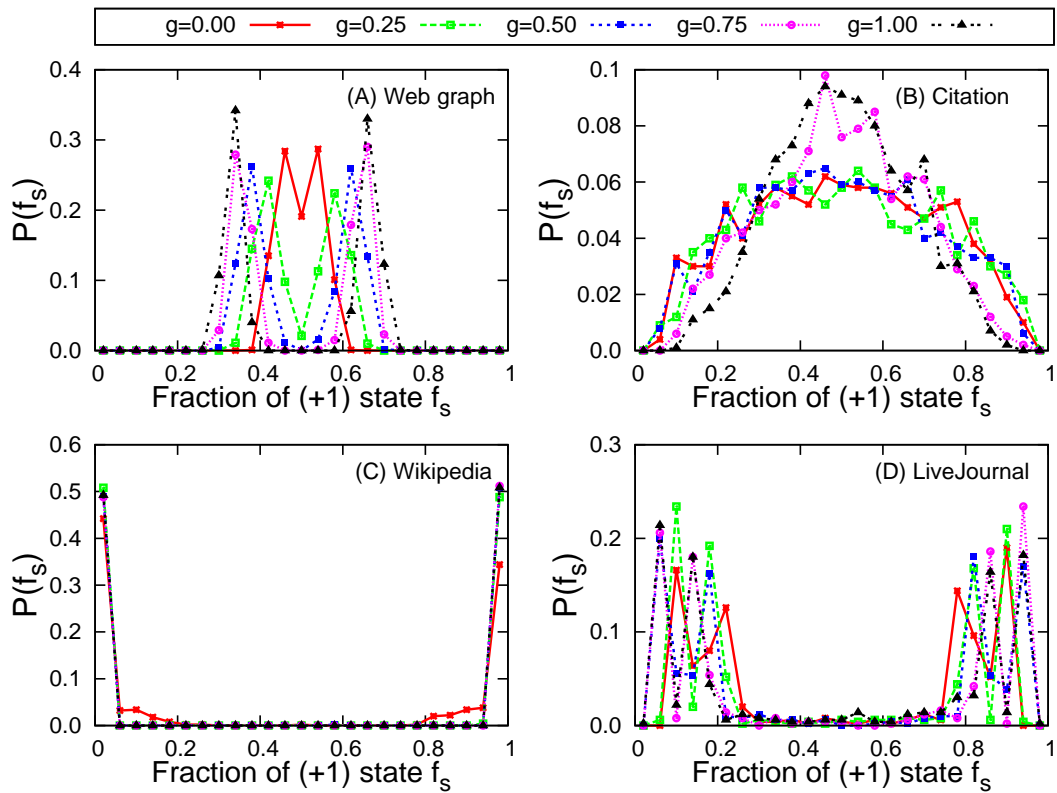}
\caption{Distributions of $(+1)$ opinion fraction in the steady state
for each empirical network. Here $f_s$ is the fraction of $(+1)$
opinion in steady state and $P(f_s)$ is the probability
distribution function of $f_s$. All the cases start with initial
fraction of $f(+1,0)=f(-1,0)=0.5$ with $1000$ realizations for Web
graph and Citation networks and $500$ realizations for Wikipedia and
LiveJournal. } \label{fig:3}
\end{center}
\end{figure*}

In order to analyze opinion formation in the steady states and
study polarization of steady state opinions, we investigate
distributions of fraction of $(+1)$ opinion $f_s$ in steady state
for each network. Fig.~\ref{fig:3} represents the distributions of
fraction of $(+1$) opinion in the steady states for each case of
empirical network starting with $f(0,+1)=f(0,-1)=0.5$. For the
cases of Web graph, Wikipedia, and LiveJournal, increasing $g$
resulted in more uniform opinion states (i.e., the fractions of
majority opinion state whether $(-1)$ or $(+1)$ are getting higher
with $g$). However, the fraction of majority opinion might not be
increasing monotonously as a function of $g$. This indicates that
a more heterogeneous node influence distribution in networks may
lead to a more ''totalitarian" society. However, the Citation
network shows the opposite pattern. It is notable that the
Citation network has different structural property from other
directed networks. Unlike the other considered networks,
reciprocal links (i.e., bi-directed links connecting from node $i$
to node $j$ and from node $j$ to $i$.) are very rare in the
citation networks due to time-ordering of citation relationships
between scientific articles (i.e., it is practically not possible
to cite publications in future). Thus this distinctive structure
might affect behaviors of collective opinion on the network.

\begin{figure*}[ht]
\begin{center}
\includegraphics[width=155mm,angle=0]{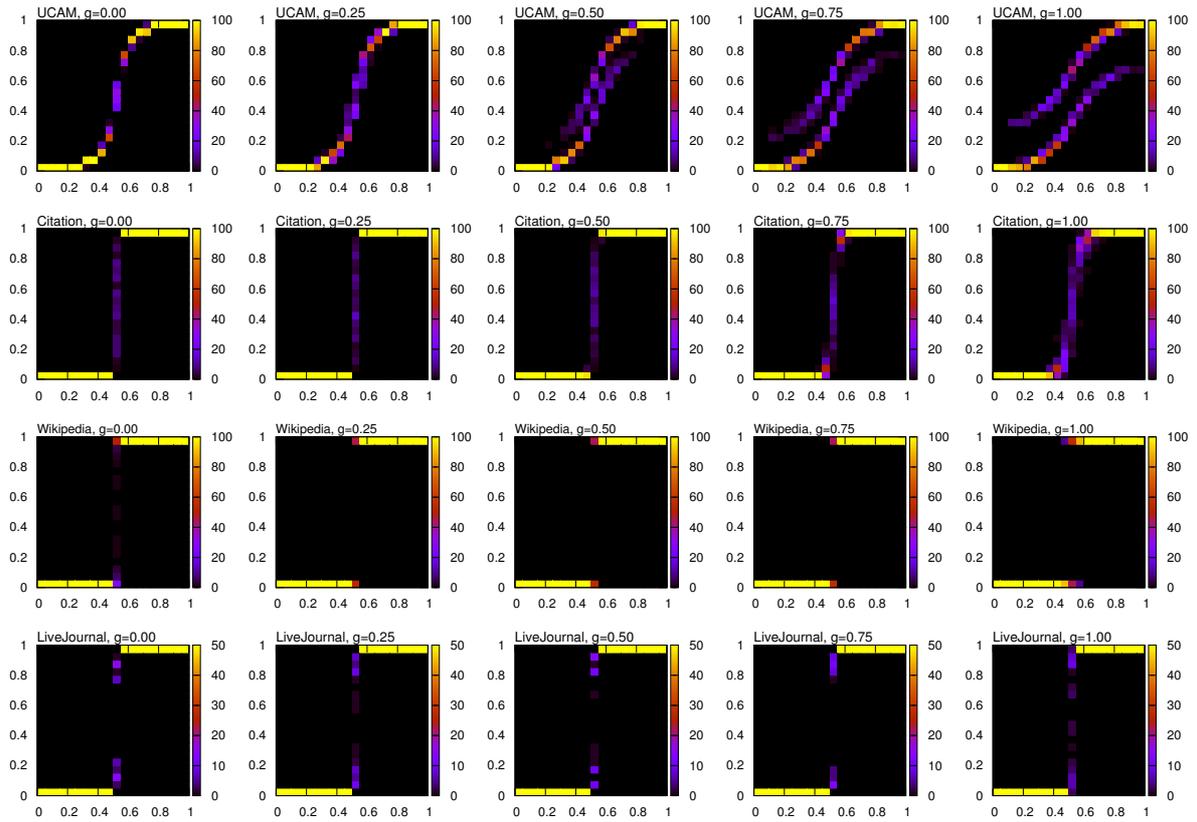}
\caption{Fraction of $(+1)$ opinion states $f_s$ ($y-$axis) in the
steady state as function of initial fraction $f_i$ ($x-$axis) of
$(+1)$ opinion state for given network and $g$. Each row
corresponds to each network and each column corresponds to the
value of $g$. Here there are $100$ realizations for Web graph,
Citation networks, and Wikipedia and  $50$ realizations for
LiveJournal. Here the color marks the relative number of cases
obtained for give values $(f_i,f_s)$, the color changes from black
(zero) to red (maximal number of cases).} \label{fig:4}
\end{center}
\end{figure*}

So far we considered only evolution of opinion states starting
from the same fractions of initial opinion states (i.e.,
$f(0,+1)=f(0,-1)=0.5$). If initial fraction of two opinions are
different, then how collective opinions on networks are formed? In
order to find out how the steady state fraction $f_s$ of nodes
with $(+1)$ opinion depends on its initial fraction $f_i=f(0,+1)$,
we investigate opinion formation with varying initial fraction of
$(+1)$ opinion and varying $g$. Fig.~\ref{fig:4} represents a
fraction of $(+1)$ opinion in the steady state $f_s$ versus an
initial fraction of $(+1)$ opinion $f_i$ for each empirical
network. Each row in Fig.~\ref{fig:4} represents each network and
each column represents each value of $g$.

In the case of Web graph, we can observe the emergence of
bistability as $g$ is increasing. Here bistability means there
exist two steady state fractions of $(+1)$ opinion. The
bistability of Web graphs is also observed in~\cite{Kandiah2012}
in the case of University of Cambridge and Oxford Web graph with
original PROF model (i.e., $g=1.0$). When $g$ is small ($g\leq
0.25$), the fraction of $(+1)$ opinion $f_s$ in the steady state
reached single value of fraction with some fluctuations.
Meanwhile, when $g\geq 0.5$, there are two values of $f_s$ in the
steady state. For LiveJournal network, there are signs of multiple
steady state fractions of $(+1)$ opinion as shown in
Fig.~\ref{fig:3}(D). This phenomenon is also observed in
Fig~\ref{fig:4} but only for $f_i=0.5$. If $f_i\neq 0.5$, we
cannot observe such multistability in the steady state. On the
other hand, there is no such bistability for the case of Citation
network and Wikipedia. In particular the Wikipedia network shows
if the initial fraction of $(+)$ opinion is less (more) than 0.45
(0.55), the final fraction is always less (more) than 0.05 (0.95).
Based on the observation, the initial fraction of the opinion
states can be critical for opinion formation in these networks but
the detail behaviors can be different depending on the types of
networks.

To characterize the effects of influential nodes on opinion
formation, we investigate how a group of selected nodes with a
fixed opinion can impose their own opinion on the entire network.
We compare two opinion implanting strategies of $n$ seed nodes
with a fixed opinion.

In the \emph{random implanting strategy}, we choose $n$ nodes as
seed nodes from a given network randomly and assign $(+1)$ opinion
to them. The opinions of seed nodes are fixed. We assign $(-1)$
opinion to the rest of nodes (i.e., non-seed nodes) in the
networks. The opinions of the non-seed nodes are flexible thus
their opinions can be changed by the modified PROF rule at each
update. Meanwhile in the \emph{targeted implating strategy}, we
choose $n$ nodes as seed nodes in order of PageRank of the nodes
and assign $(+1)$ opinion to them. The opinions of seed nodes are
also fixed. We assign $(-1)$ opinion to the rest of nodes in the
network and update the opinions of non-seed nodes by modified PROF
rule as in the random implanting strategy at each update.



\begin{figure*}[ht]
\begin{center}
\includegraphics[width=106mm,angle=-90]{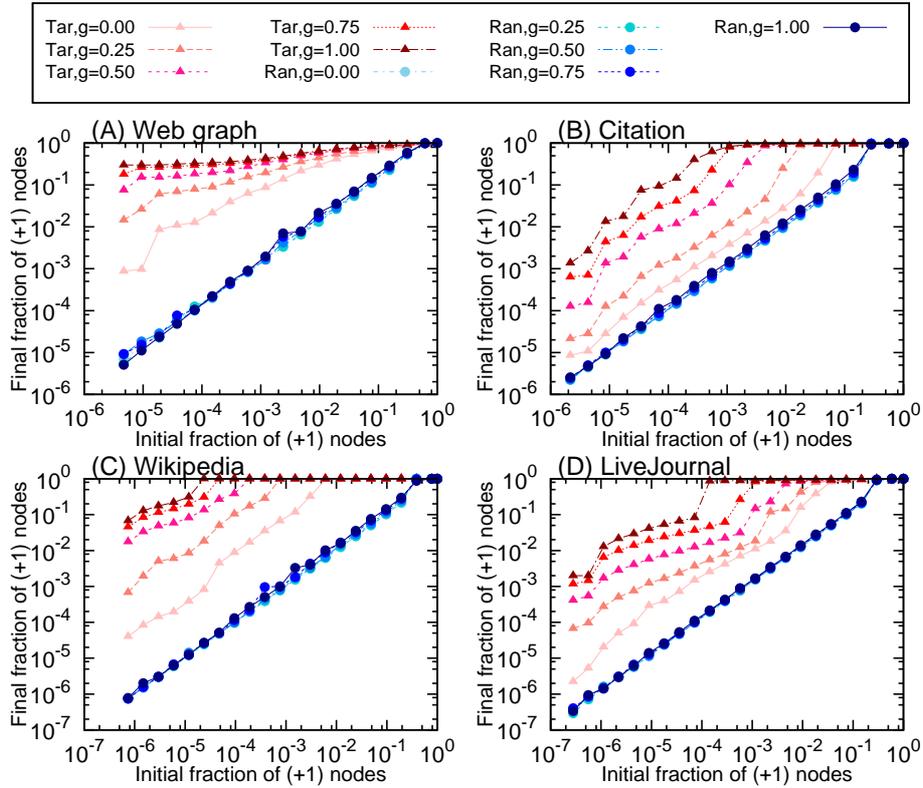}
\caption{Comparisons between the targeted implanting strategy and
random implanting strategies. ''Tar" represents the targeted
implanting strategy and ''Ran" represents the random implanting
strategy. For targeted implanting strategy (filled triangles),
pink, salmon, dark-pink, red, and dark-red colors represent
$g=0.0$, $g=0.25$,$g=0.5$,$g=0.75$, and $g=1.00$, respectively.
For random implanting strategy (filled circles), skyblue,
dark-turquoise, web-blue, blue, and navy represent $g=0.0$,
$g=0.25$,$g=0.5$,$g=0.75$, and $g=1.00$, respectively. Here there
are $100$ realizations for Web graph and Citation networks and
$50$ realizations for Wikipedia and $25$ realizations for
LiveJournal.} \label{fig:5}
\end{center}
\end{figure*}

Fig~\ref{fig:5} compares the fraction of $(+1)$ opinion nodes in
the steady state by two implanting strategies. Regardless of
networks and value of $g$, targeted implanting cases are much more
effective to lead collective opinion states of the networks to
$(+1)$ opinion. Even when $g=0.0$ (i.e., every node has the same
node influence), targeted implanting is more effective than random
implanting strategy to change the nodes in the networks to $(+1)$
opinion. The tendency is getting stronger with $g$. For the
Citation, Wikipedia, and LiveJournal networks, even a very small
fraction of top influential nodes with fixed $(+1)$ opinion (i.e.,
$f(0)\leq 0.01$) can lead to the significant fraction of $(+1)$
opinion in the steady state on the networks. For the Web graph,
the tendency is weaker partially due to the ''bistability" we
observed above. In~\cite{Kandiah2012}, it was observed that
imposing $(+1)$ opinion on small initial fraction ($\sim 1$
percent of nodes) of top PageRank nodes can lead $40$ percent of
$(+)$ opinion states. Our analysis indicates this ''elite" effect
can exist even when every node has the same influence but the
elite effect can be much stronger when node influence are
heterogeneously distributed with a larger value of $g$.

\begin{figure*}[ht]
\begin{center}
\includegraphics[width=70mm,angle=-90]{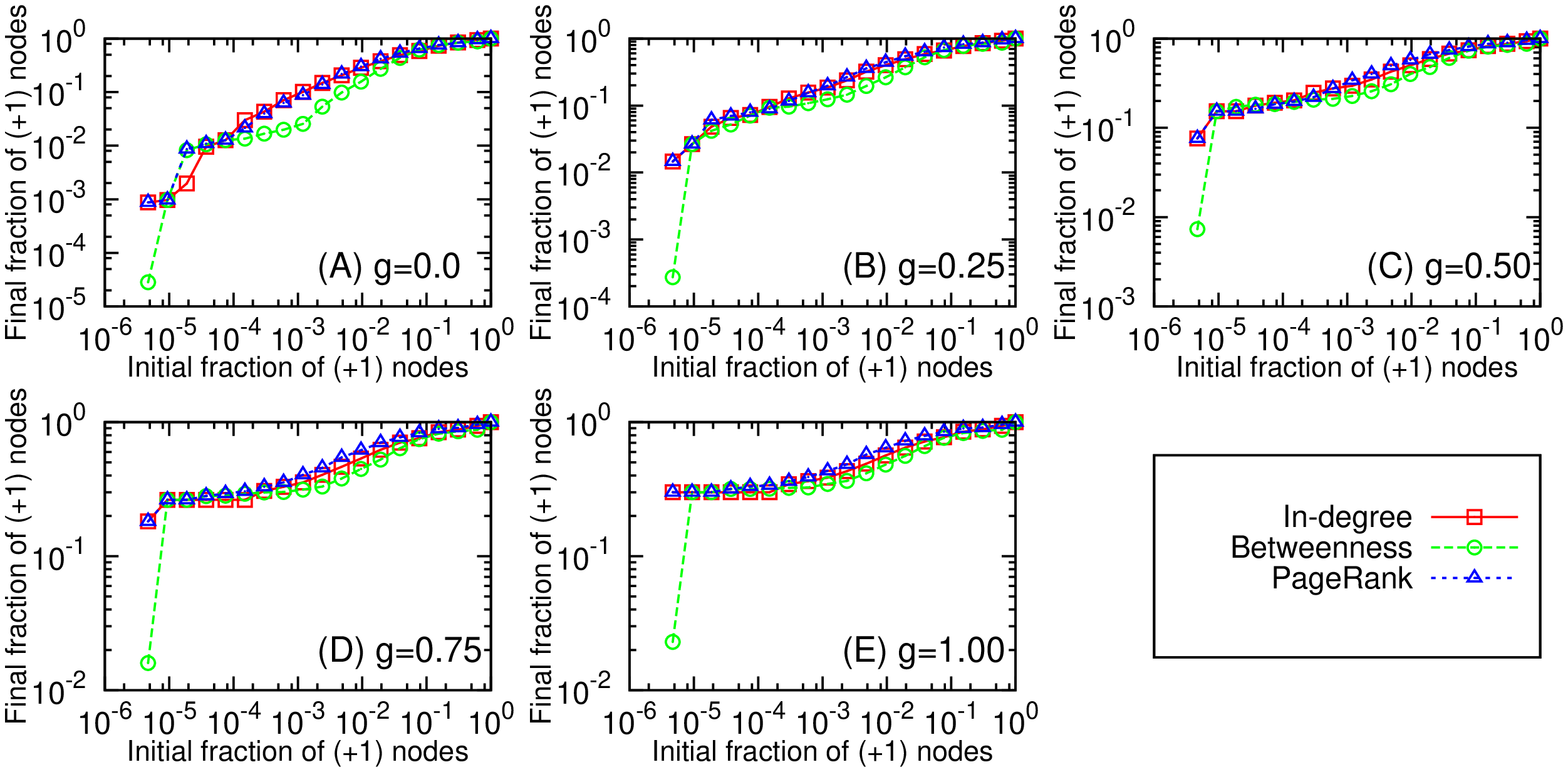}
\caption{Comparisons between the degree, betweenness, and PageRank
targeted implanting strategy. Here there are $100$ realizations
for Web graph.} \label{fig:6}
\end{center}
\end{figure*}

It would be also interesting to consider targeted implanting
strategies based on other centrality measures. We consider two
additional targeted implanting strategies based on in-degree and
betweenness centralities for Web graph since it is not feasible to
get betweenness for other networks due to their large sizes. As
shown in Fig~\ref{fig:6}, the performances of three targeted
strategies based on in-degree, betweenness, and PageRank are quite
similar with each other. We can expect similar results for the
other networks since PageRank is known to be positively correlated
with in-degree and betweenness centralities. The actual
correlation between PageRank and in-degree in Web graph is 0.886
and the correlation between PageRank and betweenness in Web graph
is 0.706.

\section{Discussion}
\label{}

Opinion formation in social systems is mediated by social
interactions between the individuals in the systems and at the
same time it is affected by influence of interacting nodes. Thus
understanding this interplay between individuals' influence and
network structure of social interactions is a salient issue. In
this study we used the modified PageRank opinion formation (PROF)
model to consider how heterogeneous node influence affects
collective opinion formation on real networks and analyzed effects
of heterogeneity of node influence on opinion formation. We found
that the relaxation time to reach the steady state is decreasing
with the heterogeneity of node influence in the networks. We also
identified that a small number of influential nodes can impose
their opinion on significant fraction of nodes, and the impacts of
these social elites on collective opinion is growing with the
heterogeneity of node influence.

All of considered networks reach a steady opinion state. However,
it is not clear why only Wikipedia shows consensus and the other
networks do not. Since we considered directed networks, asymmetric
nature of links could be the obstacle to reach consensus. To check
the effect of the asymmetric nature of links, we considered
undirected version of empirical networks but observed the same
non-consensus behaviors. Thus we can rule out this explanation. On
the other hand, a strong local structure such as communities or
modules~\cite{Girvan2002,Fortunato2010} can prohibit to reach the
consensus opinion state. Since communities in networks are typical
composed of a group of tightly connected nodes, such a densely
connected group of nodes may persist the influence from other
parts of the networks. It would be interesting to study an
interplay between influential nodes and community structure. The
Citation network also displays the opposite behaviors from the
other networks such that the other networks show more uniform
opinions states with growing $g$ while Citation network shows less
uniform steady state opinion. It will be interesting to check if
other citation networks show similar behaviors with our Citation
network.

In this study we used PageRank and its sub-linear power as node
influence. However, other node centralities on directed network
can be considered as node influence including in-degree,
betweenness centrality~\cite{Wasserman1994},
CheiRank~\cite{Chepelianskii2010}, 2DRank~\cite{Zhirov2010}, or
non-structural node attributes. Since PageRank is positively
correlated with in-degree, the study of considering node influence
which is positively correlated with in-degree can be interesting.
As described above, community  or core-periphery
structures may also significantly affect the collective opinion
formation with a local structure-based influence measure.

Due to the advent of information technology and growing usage of
social media, the problem of collective opinion formation is getting more and
more complicated going to a global scale. A quantitative understanding
of opinion formation on large-scale networks becomes of crucial importance. Our
study sheds a new light on how the node influence and network structure
together affect the collective opinion in directed networks.

\newpage

\section*{Acknowledgments}
We thank V.Kandiah for useful discussions and American Physical
Society for letting us use their citation database for Physical
Review journals. This work is supported in part EC FET Open
project ``New tools and algorithms for directed network analysis
(NADINE)" - No. 288956. Y.-H. Eom also thanks for supporting of
the EC FET project ``Financial Systems SIMulation and POLicy
Modelling (SIMPOL)" - No. 610704.





\end{document}